\documentclass[prl,floatfix,twocolumn,superscriptaddress]{revtex4}
\usepackage{times,graphicx,bbm,amsmath,amssymb}
\usepackage{epsfig,color}
\usepackage{hyperref}

\newcommand{\bmlambda}{\boldsymbol \lambda} 
\def\Tr{\hbox{Tr}} \def\sigmaCM{\boldsymbol{\sigma}}
\usepackage{pifont}

\newcommand{\ketbra}[2]{\mbox{$|#1\rangle\langle #2|$}}

\newcommand{\sab}{{\scriptstyle AB}}
\newcommand{\smb}{{\scriptstyle B}}

\newcommand{\sma}{{\scriptstyle A}}
\newcommand{\scba}{{\scriptstyle B|A}}
\newcommand{\scab}{{\scriptstyle A|B}}
\newcommand{\smx}{{\scriptstyle X}}

\def\Tr{\hbox{Tr}} \def\sigmaCM{\boldsymbol{\sigma}}
\usepackage{pifont}
\begin{document}
\title{Homodyne estimation of Gaussian quantum discord}
\author{R\'emi Blandino}
\email{remi.blandino@institutoptique.fr}
\affiliation{Laboratoire Charles Fabry, Institut d'Optique, 
CNRS, Universit\'e Paris-Sud, Campus Polytechnique, RD 128, 91127 Palaiseau cedex, France}
\author{Marco G. Genoni}
\email{m.genoni@imperial.ac.uk}
\affiliation{QOLS, Blackett Laboratory, Imperial College London, London SW7 2BW, UK}
\author{Jean Etesse}
\affiliation{Laboratoire Charles Fabry, Institut d'Optique, CNRS, 
Universit\'e Paris-Sud, Campus Polytechnique, RD 128, 91127 Palaiseau cedex, France}
\author{Marco Barbieri}
\affiliation{Laboratoire Charles Fabry, Institut d'Optique, CNRS, 
Universit\'e Paris-Sud, Campus Polytechnique, RD 128, 91127 Palaiseau cedex, France}
\author{Matteo G.A. Paris}
\affiliation{Dipartimento di Fisica, Universit\'a degli Studi di Milano,
I-20133, Milano, Italy}
\affiliation{CNISM -- Udr Milano, I-20133, Milano, Italy}
\author{Philippe Grangier} 
\affiliation{Laboratoire Charles Fabry, Institut d'Optique, 
CNRS, Universit\'e Paris-Sud, Campus Polytechnique, RD 128, 91127 Palaiseau cedex, France}
\author{Rosa Tualle-Brouri} 
\affiliation{Institut Universitaire de France, boulevard St. Michel, 75005, Paris, France}
\affiliation{Laboratoire Charles Fabry, Institut d'Optique, 
CNRS, Universit\'e Paris-Sud, Campus Polytechnique, RD 128, 91127 Palaiseau cedex, France}
\begin{abstract}
We address the experimental estimation of Gaussian quantum discord for
two-mode squeezed thermal state, and demonstrate a measurement scheme 
based on a pair of homodyne detectors assisted by Bayesian analysis
which provides nearly optimal estimation for small value of discord. 
Besides, though homodyne detection is not optimal for Gaussian discord, the noise
ratio to the ultimate quantum limit, as dictacted by the quantum
Cramer-Rao bound, is limited to about 10 dB.  
\end{abstract}
\maketitle
Quantum correlations are central resources for quantum technology. These
tight connections empower the advantages shown by the exploitation of
quantum coding in applications to cryptography, computation and sensing.
While at first entanglement was recognized to be the most peculiar form
of quantum correlations, novel concepts have been introduced to capture
either more specific aspects, such as quantum steering
\cite{wis07,sau10}, or, to the other end of the spectrum, more general
occurrences. Quantum discord represents the most successful attempt to
observe quantum features within the current picture\cite{oll01, hen02}:
it is related to the fact that quantum information in a bipartite system
can not be accessed locally without causing an inherent disturbance
– at a difference with classical probability distributions.
\par
Quantum discord has recently attracted considerable attention, due to
its possible, yet controversial, usefulness as a resource in mixed-state
quantum computing. There exist in fact architectures for which an
exponential improvement over classical resources is obtained
\cite{kni98, dat08}, albeit the entanglement becomes exponentially small
\cite{dat05, dat07}. Discord has then been suggested as the empowering
resource, while following investigations contested this interpretation
\cite{dak10}. This debate has stimulated an intense effort into looking
at protocols where discord acts a resource: it has been demonstrated
that discord does play a role in the activation of multipartite
entanglement\cite{pia11}, entanglement generation by measurement
\cite{str11}, state merging \cite{mad11}, and for complete positivity of
evolutions \cite{sha09,rod08}.  
\par
In the experimental test of such proposed connections, the comparison of
discord with relevant figures of merit is clearly connected to the
ability of estimating with the best precision allowed by a given amount
to resources.  A key problem is then to find optimal strategies, and to
understand their fundamental limit introducing proper  Cram\'er-Rao
bounds (CRB) \cite{hel67,bra94,par09}. In fact, experimental observation 
of quantu mdiscord has been
undertaken either by direct inspection of the density matrix
\cite{lan08,pao11,expD1,expD2,expD3}, or by using a witness \cite{pas11}, 
however, with
no concern about the optimality of the scheme. 
\par
Optimal estimation of quantum correlations has been investigated
for entanglement \cite{gen08} and
optimal estimators have been experimentally proved to attain the 
quantum limit for different families of qubit states \cite{bri09}. For
the perspective of quantum metrology, this is highly nontrivial, since
there exists no observable directly related to quantum discord. A proper
estimator is then needed, which might depend on several characteristic
parameters of the quantum state. In such a multiparameter problem,
finding an optimised detection scheme might be hard, and could demand
complex experimental apparata or heavy post-processing of the data.   
\par
In this Letter we demonstrate homodyne estimation of Gaussian
quantum discord in continuous variable systems \cite{gio10,ade10},
and compare the achieved level of precision with the classical
CRB for homodyne detection, and with the quantum CRB, which sets
the ultimate precision allowed by quantum mechanics. 
We found that although homodyne detection is not optimal 
for Gaussian discord, the noise
ratio to the ultimate quantum limit is limited to about 10 dB.  
Our findings
also show how a suitable Bayesian data processing may be employed to 
improve precision, especially in the estimation of small
values of discord. 
\par
Quantum discord is defined as the difference between two quantum
analogues of classically equivalent expressions of the  mutual
information in bipartite systems. Its evaluation demands an optimization
procedure over the set of all measurements on a given subsystem. For
continuous variable system, such minimisation reveals as an extremely
complex task; however, in the case of Gaussian states, we can
conveniently restrict the search to Gaussian measurement only
\cite{gio10}, obtaining
an expression for the Gaussian quantum discord \cite{gio10,ade10}. 
This sets a lower limit
to the discord of the state, and also represents an operative figure of
merit in those context where, for experimental convenience, only
Gaussian measurements are employed.  
\par
Our investigation is concerned with an important class of Gaussian
states, i.e. the two-mode squeezed thermal states (STS) naturally
produced by a non-collinear optical parametric amplifier (OPA). If we
introduce the two-mode squeezing operator $S_2(s)=\exp\left(s(a_0^\dag
a_1^\dag-a_0 a_1)\right)$, and the thermal state
$\nu(N){=}\frac{1}{N+1}\sum_n(\frac{N}{N+1})^n\ketbra{n}{n}$, we can
write the STS as \begin{align}
\label{stato}
\varrho(N_s, N_t) = S_2(s) \nu(N_t) \otimes \nu(N_t) S_2(s)^\dagger ,
\end{align}
and thus can be fully described by the two parameters $N_s{=}\sinh^2s$
and $N_t$, representing, respectively, the effective amount of {\em
squeezing photons} and {\em thermal photons}. In fact, spurious effects
such as unwanted amplification, result in a loss of purity of the
squeezed state by thermalisation, but do not affect the Gaussian
character of the emission, so the form of the density matrix
\eqref{stato} provides a fully general description of the output 
of a realistic OPA \cite{CM5}.
\par
For the class of states in Eq. (\ref{stato}) the Gaussian quantum
discord is given by 
$$
D(N_s, N_t) = h(\kappa_1) - 2 h(\kappa_2) + h(\kappa_3)\,,
$$
where $h(x)=(x+1/2)\log(x+1/2)-(x-1/2)\log(x-1/2)$ is the binary
entropy and $\kappa_1 = (1+ 2N_s)(1+2N_t)$, $\kappa_2=(N_t+1/2)$,
$\kappa_3=(1+N_s+N_t)(N_t+1/2)/(1+N_s+N_t+2 N_s N_t)$.
We can estimate the discord from $N_s$ and $N_t$ as we
varied the pump power of our OPA \cite{SM}. For each power setting,
these two parameters are extracted by the outcome of two homodyne
detectors, one on each mode, which measure pairs of quadratures
$\{X_0,X_1\}$ and  $\{P_0,P_1\}$ (Fig.1). From these, we can evaluate
the four linear combinations \begin{align}
\label{quadra}
Q^{(1/2)}= \frac{X_0 \pm X_1}{\sqrt{2}} \qquad 
Q^{(3/4)} =\frac{P_0\pm P_1}{\sqrt{2}} 
\end{align}
where $Q^{(1)}$  and $Q^{(4)}$ are squeezed quadratures, while $Q^{(2)}$
and $Q^{(3)}$ are anti-squeezed; in particular $M_q$ measurement
outcomes are recorded for each one of the four quadratures. The
corresponding variances, $\sigma^2(Q_{\rm sq})$ and $\sigma ^2(Q_{\rm
asq})$, that can be obtained from the experimental data, can be
rewritten as function of $N_s$ and $N_t$ {as follows
\begin{align}
\sigma ^2(Q_{\rm sq/asq}) &= (1 + 2 N_s \mp 2 \sqrt{N_s (1+N_s)}) (1+2 N_t)
\end{align}
}
The expressions
obtained can be then inverted to obtain the experimental estimate
$N_s^{\rm inv} $ and $N_t^{\rm inv}$, along with the relative
uncertainties $\sigma^2(N_s^{\rm inv} )$ and $\sigma^2(N_t^{\rm inv} )$.
These values can be used in the expression for discord \cite{SM} to
calculate its value $D^{\rm inv}$, and the uncertainty
$\sigma^2(D^{\rm inv})$. 
The uncertainties on these quantities are then obtained by a Monte Carlo 
procedure \cite{SM}.
\begin{figure}
\includegraphics[width=0.9\columnwidth, viewport= 140 220 650 500, clip]{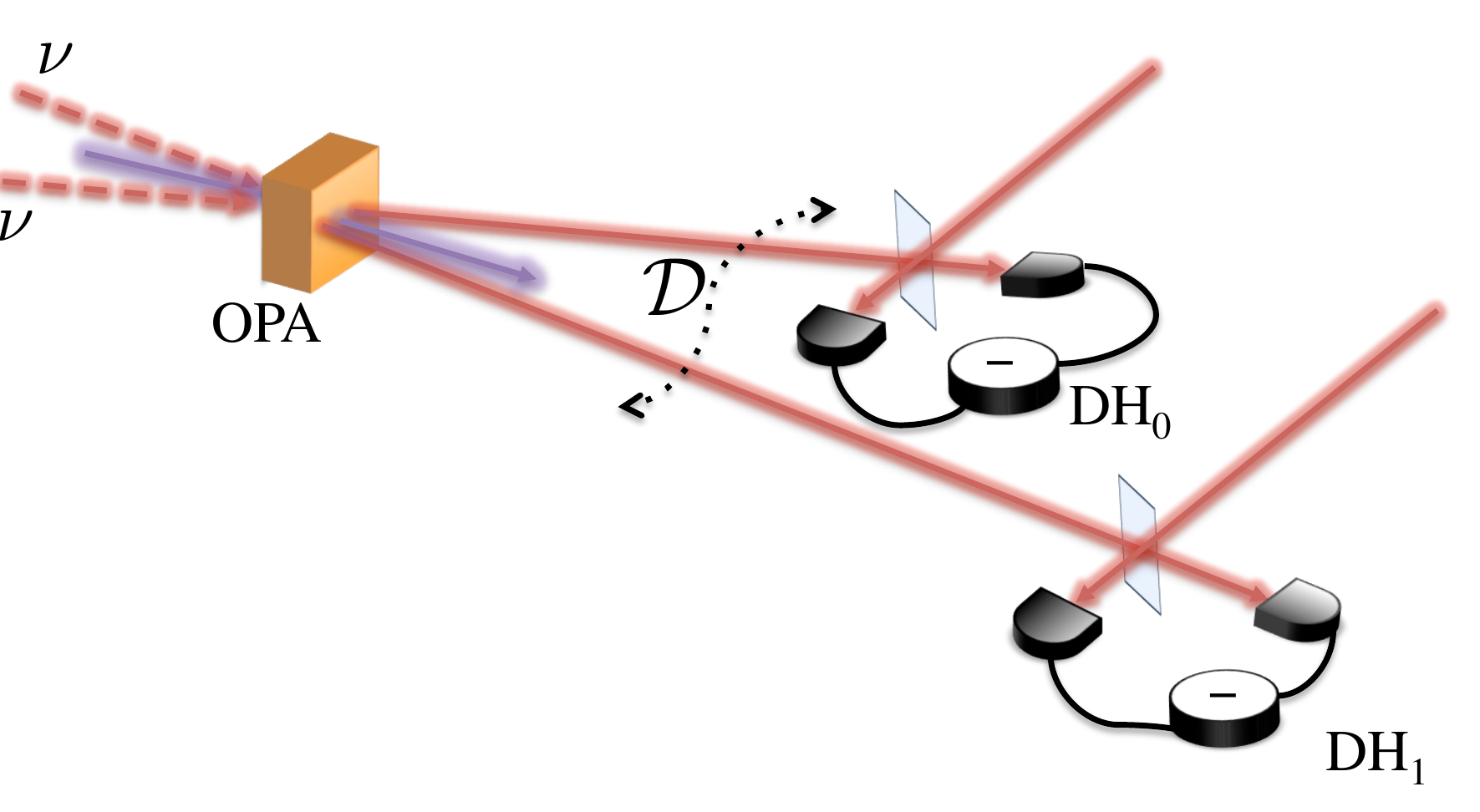}
\caption{Conceptual layout of the discord estimation. Our non-collinear
OPA is based on a nonlinear KNbO$_3$ cristal, pumped by a
frequency-doubled Ti:Sapph laser (repetition rate 800kHz, wavelength
$\lambda_p{=}425$nm, pulse duration 120fs). This produces a STS with
discord $D$ depending on the pump power, i.e. on their average
photon number. the two mode are measured by two homodyne detectors
DH$_0$ and DH$_1$. The relative phase between the local oscillators is
locked by mapping them to different polarisations on the same spatial
mode. In this way, we can record blocks of 20000 values of pairs of
quadratures for $\{X_0,X_1\}$ and the same for $\{P_0,P_1\}$.} \label{fig:setup}
\end{figure}
One can use the same data and refine the estimation by using a Bayesian
analysis. As described above, each data sample corresponds to $M_q=2\cdot 10^4$
measurement of each of the four quadratures. The total sample, thus
correspond to $M_T=4 M_q$ homodyne outcomes 
\begin{align}
\cal{X}&=\{q^{(1)}_1,.., q^{(1)}_{M_q}, q^{(2)}_1,.., q^{(2)}_{M_q},
q^{(3)}_{1},.., q^{(3)}_{M_q}, q^{(4)}_1, .. , q^{(4)}_{M_q} \}.
\notag
\end{align}
The overall sample probability can be evaluated as 
\begin{align}
p({\cal{X}} | N_s , N_t ) = \prod_{k=1}^4 \prod_{j=1}^{M_q} p_{k} (q^{(k)}_j | N_s, N_t) 
\end{align}
where the probability of obtaining the outcome $q^{(k)}_j$ by measuring the quadrature
$Q^{(k)}$ is a Gaussian distribution 
\begin{align}
\label{apriori}
p_{k} (q^{(k)}_j | N_s, N_t)  = \frac{1}{\sqrt{2 \pi \sigma_k^2}} \exp
\left( - \frac{(q^{(k)}_j)^2}{2 \sigma_k^2} \right).
\end{align} 
For squeezed quadratures ($k= \{ 1, 4\}$)  we substitute $\sigma_k^2 =\sigma^2 (Q_{\rm sq})$,
while for anti-squeezed quadrature ($k= \{ 2, 3 \}$) $\sigma_k^2 =\sigma^2 (Q_{\rm asq})$.
By means of the Bayes theorem, we obtain the {\em a-posteriori} probability
\begin{align}
&p( N_s, N_t | {\cal X}) = \frac{1}{\mathcal{N}}\, p({\cal X} | N_s, N_t) p_0(N_s) p_0
(N_t), \label{eq:posteriori} \\ &\mathcal{N} = \int dN_s \:dN_t \: p({\cal X} |
N_s, N_t) p_0(N_s) p_0 (N_t).  \end{align}
where the $p_0(N_s)$ and $p_0(N_t)$ are the so-called {\em a-priori} probability 
distributions for the two parameters. 
In our procedure, we use the results of the inversion estimation to
construct these {\em a-priori} distributions. That is, we consider $p_0(N_s)$ and 
$p_0(N_t)$ as Gaussian functions with respectively, mean values equal to 
$N_s^{\rm inv}$ and $N_t^{\rm inv}$, and variances equal to $\sigma^2(N_s^{\rm inv})$
and $\sigma^2 (N_t^{\rm inv})$.
Then, we can use the {\em a-posteriori} probability distribution evaluated as in 
Eq. (\ref{eq:posteriori}) to obtain an estimate of the two parameters
and of their variances. In formula, (for $j=s,t$)
\begin{align}
N_j^{\rm bay} &= \int dN_s \: dN_t \: N_j \: p( N_s, N_t | {\cal X}) \\
\sigma^2(N_j^{\rm bay}) &= \int dN_s \: dN_t \: (N_j - N_j^{\rm bay})^2 
\: p( N_s, N_t | {\cal X})
\,.
\end{align}
By using the formula of the discord for two-mode STS in \cite{SM}
and by propagating the errors, we then obtain an estimate $D^{\rm bay}$
for the discord, alongs with its variance $\sigma^2(D^{\rm bay})$.
\par
The value of discord depends on both the squeezing and thermal photons.
Consequently, its estimation is inherently a multi-parameter problem,
and we have to identify the relevant physical parameters to evaluate the
correct CRB. In the multiparametre scenario, the quantum Fisher
information (QFI) associated to a vector of parameters $\bar
\lambda{=}\{\lambda_i\}_{0\leq i\leq n}$ is in the form of a matrix
$\boldsymbol{H}$. This sets a lower bound on the covariance
$\sigma^2_{ij}{=}\langle \lambda_i \lambda_j\rangle{-}\langle \lambda_i
\rangle\langle\lambda_j\rangle$ after $M$ repetitions on the experiment:
\begin{equation} 
\label{bound} \sigma^2_{ij}\geq \frac{1}{M}
\left({\boldsymbol{H}}^{-1} \right)_{ij}
\end{equation} 
In the specific case of our
experiment, we can bound the uncertainty on the discord $D$ of the
states we prepare as: $\sigma^2({D})\geq \frac{1}{M}
({\boldsymbol{H}}^{-1})_{DD}$. While our measurement strategy has the
advantage of being simple, it is not expected to be optimal, i.e. to 
saturate the quantum CRB. In order to assess the estimator, i.e. the data 
processing, we also need to compare it to the classical CRB associated to our
specific measurement, which is analogously described by a classical 
Fisher information (FI) matrix $\boldsymbol{F}$. 
\par
In the evaluation of the correct bound, we need a suitable
parametrisation of the state, so that in the expression \eqref{bound}
one parameter only actually varies, while the others are kept fixed:
this can not be the case for the number of thermal and squeeing photons,
as both of them change with the pump power. Therefore, we need to
reshape the QFI matrices for different couples of parameters, so to
consider those which are more directly connected to the experimental
conditions.  We start by considering the first couple ${\bmlambda}_1 = \{
N_s, N_t\}$; by using the formulas described in \cite{SM}, we obtain
\begin{align}
\label{H1}
\boldsymbol{H}^{(1)} = {\rm diag} \left( \frac{(1+2 N_t)^2}{N_s(1+N_s)(1+2 N_t + 2 N_t^2)}, 
\frac{1}{N_t (1+N_t} \right)\,.
\end{align}
As explained above, thermal photons appear because of imperfections in
the operation of the OPA and because of loss. When the squeezing is not
too low, we can reparametrise
our state by taking in consideration the effective squeezing strength
$r$, and a parasite amplification with strength $\gamma r$ \cite{bru09,
bar10}. The overall homodyne detection 
can be separately calibrated, obtaining $\eta=0.62$. Thus we can rewrite
the matrix \eqref{H1} in terms of the two unknown physical parameters 
${\bmlambda}_2 = \{r, \gamma\}$ via the
expression $\boldsymbol{H}^{(2)} = B_{12} \boldsymbol{H}^{(1)}
B_{12}^T$, where $B_{12}$ is the transfer matrix for this change of
variables \cite{SM}.  Next, since the physical parameter that changes during our
experiment, resulting in the variation of the amount of discord, is the
squeezing parameter $r$ (while $\gamma$ and $\eta$ can be considered to
remain constant), we perform the last change of variable, by considering
${\bmlambda}_3 = \{ D , \gamma \}$. Again the QFI matrix can be obtained as
$\boldsymbol{H}^{(3)} = B_{23} \boldsymbol{H}^{(2)} B_{23}^T$, 
and
the bound on the variance for the quantum discord can be easily evaluated
as described in Eq. (\ref{bound}).
\par
\begin{figure}
\includegraphics[width=0.9\columnwidth]{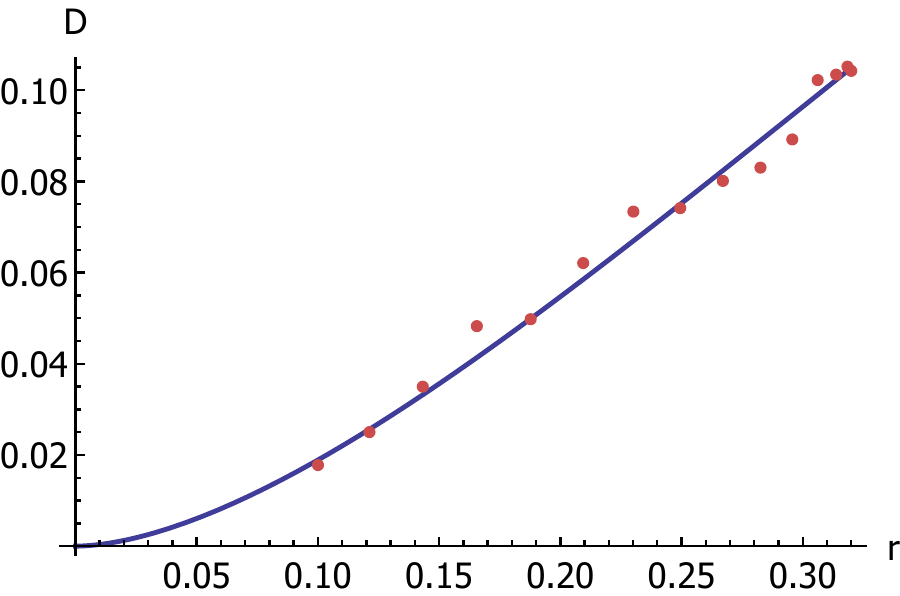}
\caption{Experimental values of Gaussian quantum discord from homodyne
data and Bayesian estimation. The points correspond to the estimated 
experimental values, while the solid line is the theoretical prediction
for $\eta{=}0.62$ and $\gamma{=}0.73$ (the value of $\gamma$ has been
extracted from a best-fit of the points). 
Uncertainties are within the point size.}
\label{fig:discord}
\end{figure}

We also want to derive the classical CRB for quantum
discord, that we obtain if we consider as measurement homodyne detection
of squeezed and anti-squeezed quadratures of a two-mode squeezed thermal
state. Let us start by considering the Fisher information matrix we
obtain if we want to estimate the two parameters $\bmlambda_1 = \{N_s, N_t
\}$ by means of homodyne detection on a certain quadrature $Q_{\phi}$.
Since the state is a Gaussian state, the conditional probability
distribution of measuring a value $x$, is a Gaussian function, with zero
mean, and variance $\sigmaCM^2(Q_\phi)$. By using the formulas in
\cite{SM} and evaluaiting some Gaussian integrals, one easily obtains
the following formula for the Fisher matrix elements \begin{align}
\label{CCR}
{\boldsymbol F}_{\mu\nu} = \frac{1}{2 \sigma^2(Q_\phi)} \frac{\partial \sigma^2(Q_\phi)}
{\partial \lambda_\mu}
\frac{\partial \sigma^2(Q_\phi)}{\partial \lambda_\nu}  
\end{align}
where $\lambda_{\mu}=\{ N_s, N_t\}$.  
If one considers to measure the squeezed or the anti-squeezed quadratures
one obtains the following FI matrices:
\begin{align}
\boldsymbol{F}^{\rm sq/asq} &= \left(
\begin{array}{ c c}
\frac{1}{2 N_s + 2N_s^2} & \mp \frac{1}{\sqrt{N_s (1+N_s)}(1+2N_t)} \\
 \mp \frac{1}{\sqrt{N_s (1+N_s)}(1+2N_t)} & \frac{2}{(1+2 N_t)^2} 
\end{array}
\right) \notag 
\end{align}
If we perform a fixed number of measurements, where half of them are done on the squeezed
quadratures, and the remaining ones on the anti-squeezed quadratures, 
the overall FI matrix which will give the CRB for the two
parameters $\bmlambda_1 = \{ N_s, N_t\}$ is obtained as
\begin{align}
\boldsymbol{F}^{(1)} =  \frac12 (\boldsymbol{F}^{\rm sq} 
+ \boldsymbol{F}^{\rm asq})=
%
\hbox{diag}(
\frac{1}{2 N_s + 2N_s^2}, \frac{2}{(1+2 N_t)^2 })
\end{align}
To obtain the CRB for homodyne detection of Gaussian discord, we can proceed as we showed for the
quantum CRB, simply replacing the QFI matrices, with the FI ones.
The values of the discord obtained using our Bayesian estimation are
shown in Fig. \ref{fig:discord}: the points indicate the experimental
data, while the solid line describes the model \eqref{stato}, where the
homodyne efficiency $\eta$ and the relative parasite gain $\gamma$ are
kept to a constant value. Our model is in satisfactory agreement with the data,
so we can be confident of that the CRB calculated after the matrix
\eqref{H1} reliably describes the ultimate limit for precision.  
In the formulae of the Bayes rule \eqref{eq:posteriori}, we need to
multiply several probabilities \eqref{apriori}, which rapidly give a
number hardly manageable by reasonable computing power: this sets a
limit to the number of quadrature values one can  effectively use in
about 800 points. In order to use larger samples, we have divided our
data in $N_b=10^2$ blocks of 200 points for each of the quadratures
\eqref{quadra}, calculated the Bayesian estimation of the discord for
each block, then considered the average weighted on the associated
uncertainties. We notice that the {\it a priori} probabilities
\eqref{apriori} are calculated from the whole set of data containing
$M_T$ values: as they intervene in the evaluation for each block, the
overall number of resources to be considered is $M = N_b\cdot M_T$.
\par
\begin{figure}
\includegraphics[width=0.9\columnwidth]{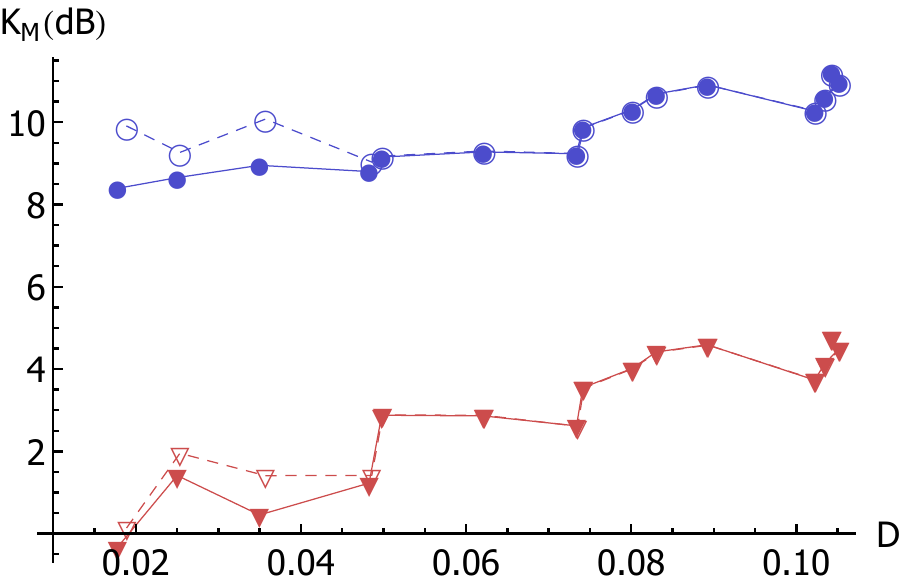}
\caption{The noise ratio $K_M$ as a function of discord. Circles and triangles 
correspond respectively to the quantum and the classical CRB. Solid points refer to Bayesian
estimation, while empty ones correspond to estimation by inversion. The uncertainties for the inversion method are estimated using a MonteCarlo procedure on $10^6$ points. Notice that
the number of resources for Bayesian estimation is $M=N_b \cdot M_T$, 
while for the inversion method $M=M_T$. 
}
\label{fig:CCR}
\end{figure}

The comparison between our experimental uncertainties and both the 
Cram\'er-Rao limit for our detection \eqref{CCR} and the quantum 
Cram\'er-Rao limit 
is shown in Fig.\ref{fig:CCR}, where we report the quantity
$K_M = M \sigma^2 (D)/({\boldsymbol F}^{-1})_{DD}$(or the analogue
quantity involving the QFI) expressed in dB.
$K_M$ is the variance of the discord estimator from homodyne data 
multiplied by the number of resources and divided by the relevant
elements of the (quantum) inverse Fisher matrix. For $K_M$ equal to
unity we have optimal estimation. Solid points refer to Bayesian
estimation while empty ones correspond to estimation by inversion.
We notice that for low values of discord, 
the Bayesian technique provides a nearly optimal estimator for the chosen
measurement strategy, whereas estimation by 
inversion is  noisier. We also notice that the point corresponding to 
the lowest value of the discord is slightly below the quantum CRB: this 
confirms that for low values of
the squeezing of the pump, the model we use is not as accurate as in 
other regimes.
For
increasing values, the observed variances depart from the optimum by
less than an order of magnitude: as Bayesian estimation rapidly
converges to optimal, we can attribute this trend to actual variations
of the value of the discord in the experiment, becoming more important
than statistical fluctuations when the discord increases.
\par
The measurement we have adopted has the considerable advantage of being
the simplest experimental option; however, simplicity always comes at a
price, and we do not expect it to deliver the best estimator for discord
as established by the quantum CRB.  In the limit
of low discord, we measure a ratio of about 10 dB , which tells us that
the price we have to pay is quite reasonable. The departure from the
quantum CRB then slightly increases with discord.
\par
In conclusion, we have presented the experimental estimation of Gaussian
quantum discord for two-mode squeezed state. Our scheme is based on
homodyne detection assisted by Bayesian analysis.  Our results are in
good agreement with the theorerical model, and this allows us to perform
a reliable precision analysis. We found that homodyne estimation shows
about 10 dB of added noise compared to the ultimate bound imposed by the
quantum Fisher information, with Bayesian analysis that slighlty
improves performances for small values of discord.  We have also
compared our results with the CRB for homodyne detection and found that
the estimation is nearly optimal for small values of discord.  The
\par
usefulness of quantum discord as a resource for quantum technology is a
heavily debated topic, and a definitive answer  may only come from
experiments involving carefully prepared quantum states. Our results
contribute to the precise characterization of Gaussian discord and
illustrate how a suitable data processing may decrease the uncertainty
when optimal detection schemes are not available.
\par
We thank A. Datta, G. Adesso, S. Olivares and M. Paternostro
for discussion and comments. We acknowledge support
from the EU project COMPAS, the 
MIUR project FIRB
LiCHIS-RBFR10YQ3H,  
and the ERANET project HIPERCOM.
MGG acknowledges a fellowship support from UK EPSRC (grant
EP/I026436/1), MB is supported by the Marie Curie contract
PIEF-GA-2009-236345-PROMETEO.

\vfill\newpage
\widetext
\section{Supplemental Material}
\subsection{Definition of quantum discord}
The state of a bipartite system $\varrho_{\sab}$ is called separable if it can be produced by local operations and classical communication, viz. $\varrho_\sab = \sum p_k \sigma_{\sma k}\otimes\sigma_{\smb k}$, $\sigma_{\sma k}$ and $\sigma_{\smb k}$ are generic density matrices describing the states of the two subsystems. Despite the fact that uniquely classical information is exchanged, this procedure can neverless generate quantum correlations, as revealed by inspecting the mutual information of the two subsystems. This is the idea on which quantum discord is built.
\par
For classical variables, mutual information can be defined by the equivalent expressions
\begin{equation}
\label{classical}
I(A;B)=H(A)+ H(B) - H(A,B),
\end{equation}
and 
\begin{equation}
\label{altra}
I(A;B) = H(A) - H(A|B) \equiv H(B) - H(B|A),
\end{equation}
where $H(X)= - \sum_x p_\smx (x)\, \log p_\smx (x)$ is the Shannon entropy
of the corresponding probability distribution and the 
conditional entropy is defined as $H(A|B)=-\sum_b p_\smb (b) \sum_a 
p_\scab (a|b)\,\log p_\scab (a|b)=-\sum_{ab} p_\sab (a,b)\, 
\log p_\scab (a|b)$, where we used the joint 
probability $p_{\sab}(a,b)$, the two marginal probabilities 
$p_\sma (a) \equiv \sum_b p_{\sab}(a,b)$ and $p_\smb (b)
\equiv \sum_a p_{\sab}(a,b)$, and the conditional probabilities
$p_\scab (a|b)= p(a,b)/p(b)$ and $p_\scba (b|a) = 
p(a,b)/p(a)$.

The idea of quantum discord grows out of the fact that the 
quantum version of the mutual information of a bipartite state 
$\varrho_\sab$ may be defined in two inequivalent ways. The first is 
obtained by the straightforward quantization of the classical expression \eqref{classical}, 
\begin{equation}
I(\varrho_\sab)=S(\varrho_\sma) + S(\varrho_\smb) - S(\varrho_\sab)
\end{equation}
where $S(\varrho)= - 
\hbox{Tr}[\varrho\,\log\varrho]$ is the Von-Neumann entropy of 
the state $\varrho$ and 
$\varrho_\sma=\hbox{Tr}_\smb[\varrho_\sab]$,
$\varrho_\smb=\hbox{Tr}_\sma[\varrho_\sab]$ 
are the partial traces over the two subsystems.
On the other hand, the quantization of the expression based
on conditional entropy involves the conditional state of a 
subsystem after a measurement performed on the other one.
This fact has three relevant consequences: 
\begin{itemize}
\item the symmetry between the two subsystems is broken; 
\item this quantity depends on the choice of the measurement;  
\item the resulting expression is generally different from $I(\varrho_\sab)$. 
\end{itemize}
Let us denote by $\varrho_{\sma k} = \hbox{Tr}_\smb 
[\varrho_\sab\, {\mathbb I}\otimes P_k]/p_\smb (k)$ with $p_\smb (k)=\hbox{Tr}_\sab 
[\varrho_\sab\, {\mathbb I}\otimes P_k]$, the state of the
system $A$ conditioned on the outcome $k$ from a measurement
performed on the system $B$; $\{P_k\}$ denotes the elements of a POVM .
The quantum analogue of the expression  \eqref{altra}
 is then defined as the upper bound
\begin{equation} 
\label{jquantum}
J_\sma = \sup_{\{P_k\}} S(\varrho_\sma) - \sum_k p_\smb (k) 
S(\varrho_{\sma k})
\end{equation}
taken over all the possibile measurements. This represents the maximal Holevo information that can be achieved by using the subsystem $A$ while adopting subsystem $B$ as a measuring device for $A$.

Finally, the quantum A-discord is defined in terms of the 
mismatch $ D(\varrho_\sab) = I(\varrho_\sab) - 
J_\sma(\varrho_\sab)$. Analogously one is led to define 
the B-discord through the entropy of conditional states of 
system $B$. 

The direct transposition of these definitions to the continuous variable realm is hindered by the complexy of the maximisation requested in \eqref{jquantum} in an infinite-dimensional space. For Gaussian states, one can introduce the notion of Gaussian quantum discord by restricting the maximisation to Gaussian POVMs, which lead to an analytical expression. In order to obtain an explicit expression, we write the covariance matrix of the state in the form
\begin{equation}
\Sigma{=}\left(
\begin{array}{cccc}
a&0&c&0\\
0&a&0&-c\\
c&0&b&0\\
0&-c&0&b
\end{array}
\right).
\end{equation}
The symplectic invariants are then given by $I_1{=}a^2$, $I_2{=}b^2$, $I_3{=}-c^2$, $I_4=\det(\Sigma)$. We can calculate the A-discord as
\begin{equation}
{D}(\Sigma){=}h(\sqrt{I_2})-h(d_-)-h(d_+)+h\left(\frac{\sqrt{I_1}+2\sqrt{I_1I_2}+2I_3}{1+2\sqrt{I_2}}\right).
\end{equation}
In the formula above we have introduced the symplectic eigenvalues $d^2_\pm{=}\frac{1}{2} \left[ \Delta\pm\sqrt{\Delta^2{-}4I_4} \right]$, $\Delta{=}I_1+I+2+2I_3$, and $h(x)$ is the binary entropy
\begin{equation}
h(x)=(x+1/2)\log(x+1/2)-(x-1/2)\log(x-1/2)\end{equation}

For symmetric states - such as those considered in the present investigation - the distinction between A-discord and B-discord becomes superfluous and one can simply talk about the discord of the state.

\subsection{Multiparametric quantum estimation} 
Here we present the case where the estimation of more than one parameter
has to be performed. We define a family of quantum states
$\varrho_{\bmlambda}$ which depends on a set of $N$ parameters 
$\bmlambda=\{\lambda_\mu \}$, $\mu=1,\dots,N$. 
In this case the geometry of the
estimation problem is contained in the QFI matrix, whose elements are
defined as 
\begin{align}
{\boldsymbol H}(\bmlambda)_{\mu\nu} 
&=\Tr\left[\varrho_{\bmlambda} \frac{ L_\mu L_\nu  + L_\nu L_\mu}{2} \right],
\end{align}
and where we have introduce the Symmetric Logarithmic Derivatives (SLD) $L_\mu$ 
corresponding to the parameter $\lambda_\mu$, as the selfadjoint operator that 
satisfies the equation
\begin{align}
\frac{L_\mu \varrho_{\bmlambda} + \varrho_{\bmlambda} L_\mu}{2} = 
\frac{\partial\varrho_{\bmlambda}}{\partial \lambda_\mu}.
\end{align}
In terms of eigenvalues and eigenvectors of $\varrho_{\bmlambda}$, by denoting
with $\partial_\mu$ the partial derivative respect to $\lambda_\mu$, we have
\begin{align}
{\boldsymbol H}(\bmlambda)_{\mu\nu}&= \sum_n \frac{(\partial_\mu a_n) (\partial_\nu a_n)}{a_n} + 
\sum_{n\neq m}
\frac{(a_n-a_m)^2}{a_n + a_m} \times \nonumber \\
&\times (\langle \psi_n|\partial_\mu \psi_m\rangle
\langle \partial_\nu \psi_m| \psi_n \rangle +
\langle \psi_n|\partial_\nu \psi_m\rangle
\langle \partial_\mu \psi_m| \psi_n \rangle ).
\end{align}
The QFI matrix here defined provides a lower bound (the quantum Cram\'er-Rao bound)
on the covariance matrix ${\boldsymbol \gamma}_{\mu\nu}= 
\langle \lambda_\mu \lambda_\nu \rangle - \langle \lambda_\mu\rangle\langle
\lambda_\nu\rangle$, {\em i.e.},
\begin{align}
{\boldsymbol \gamma} 
\geq \frac{1}{M}{\boldsymbol H}(\bmlambda)^{-1}.
\end{align}
In the multiparametric case this bound is not in general achievable, on the other
hand, the diagonal elements of the inverse Fisher matrix provide
achievable bounds for the variances of single parameter estimators, 
at fixed value of the others
\begin{align}
{\mathrm{Var}}(\lambda_\mu) = {\boldsymbol \gamma}_{\mu\mu} 
\geq \frac{1}{M}{\boldsymbol H}(\bmlambda)^{-1}_{\mu\mu}.
\label{eq:QCRMulti}
\end{align}
Let us now suppose that we are interested in the estimation of 
different set of parameters  $\tilde{\bmlambda}= \{\tilde{\lambda}_\nu=
\tilde{\lambda}_\nu(\bmlambda) \}$ which are functions of the 
previous ones. We then need to reparametrize the family of quantum states 
in terms of $\tilde{\bmlambda}$.
Since $\tilde{\partial}_\nu = \sum_\mu B_{\mu\nu} \partial_\mu$ with $B_{\mu\nu} =
\partial \lambda_\mu/\partial \tilde{\lambda}_\nu$ we have that
\begin{align}
\widetilde{L}_\nu = \sum_\mu B_{\mu\nu} L_\mu
\end{align}
and the new QFI matrix simply reads
\begin{align} 
\widetilde{{\boldsymbol H}}=B
{\boldsymbol H}{B}^T. \label{eq:Hchangevar}
\end{align}
\par
We consider here the case where we perform a specific indirect measurement
in order to infer the values of the parameters $\bmlambda$, given some measurement 
outcomes ${\cal X}=\{x_1,x_2,\dots \}$. 
The whole measurement process can be described by the conditional probability
$p(x|\bmlambda)$ of obtaining the value
$x$ from the measurement when the parameters have the values  $\bmlambda$. 
Given this object, we can define  the Fisher information (FI) matrix whose elements are obtained as
\begin{align}
{\boldsymbol F}_{\mu\nu} = \int dx \: p(x|\bmlambda) 
\frac{\partial \ln p(x|\bmlambda)}{\partial \lambda_\mu}
\frac{\partial \ln p(x|\bmlambda)}{\partial \lambda_\nu}.
\end{align}
This matrix defines a bound on the covariance matrix ${\boldsymbol \gamma}$
for the specific measurement we performed. 
In particular we are interested in the bound for the variance of 
a single-parameter, at fixed values of the others, which reads
\begin{align}
{\mathrm{Var}}(\lambda_\mu) 
\geq \frac{1}{M}{\boldsymbol F}(\bmlambda)^{-1}_{\mu\mu} \geq
\frac{1}{M}{\boldsymbol H}(\bmlambda)^{-1}_{\mu\mu} 
\end{align}
and which in turn is always lower bounded by quantum CRB given 
in Eq. (\ref{eq:QCRMulti}).\\
Notice that if we have to reparametrize our family of states in terms of different parameters 
$\widetilde{\bmlambda}$, we can use the same formulas shown above for the QFI matrix,
obtaining the new FI matrix as $\widetilde{\boldsymbol F} = B {\boldsymbol F}B$.
\subsection{Physical model and evaluation of quantum discord}
Here we give explicit expressions of the formulas used in the main text. 
A two-mode squeezed thermal state (STS) is fully characterized by 
the two parameters $N_s = \sinh^2s$ and $N_t$, 
representing, respectively, the effective amount of squeezing photons and 
thermal photons. In our experimental model, 
these quantities can be obtained as a function of the physical parameters
$\{ r, \gamma, \eta\}$, that is
\begin{align}
N_s &= \frac{1}{2} \left( 
-1 + \frac{A(r,\gamma,\eta)}
{\sqrt{\eta^2 \cosh^4{r} \cosh^2(2r\gamma) + B(r,\eta)^2 +
2\eta\cosh^2 r (-2\eta\cosh^4 (r\gamma) \sinh^2 r + \cosh(2r\gamma) B(r,\eta))}}
\right) \nonumber \\
N_t &= \frac{1}{2} \left(
-1 + \sqrt{(A(r,\gamma,\eta) - \eta\cosh^2(r\gamma)\sinh{2r} )
(A(r,\gamma,\eta)+ \eta\cosh^2(r\gamma)\sinh{2r} )}
\right) \label{eq:NsNt}
\end{align}
where
\begin{align}
A(r,\gamma,\eta) &= 1-\eta+\eta \cosh^2 r \cosh{2r\gamma} + \eta\sinh^2 r \\
B(r,\eta) &= 1-\eta+\eta\sinh^2 r .
\end{align}
Notice that by varying the pump power, we change the parameter $r$ only, while the
noise parameters $\gamma$ and $\eta$ stay constant to a very good level of 
approximation (it should
depend on the mode matching only). As a result, both the effective squeezing and thermal
photons $N_s$ and $N_t$ change accordingly. 
\\
The covariance matrix of a two-mode STS can be written as
\begin{align}
\Sigma_{sts} = \left(
\begin{array}{ c c }
a\mathbbm{1}_2 & c\: \sigma_z \\
c\: \sigma_z & a \mathbbm{1}_2
\end{array}
\right)
\end{align}
where
\begin{align}
a &= (1+2 N_t) (1+2 N_s)  \\
c &= 2 (1+2 N_t)\sqrt{N_s(N_s+1)},
\end{align}
and $\mathbbm{1}_2$ and $\sigma_z$ are respectively the
$2\times 2$ identity matrix and the Pauli matrix for the $z$ direction.
Following \cite{gio10}, the quantum discord can be thus evaluated, obtaining
\begin{equation}
\begin{aligned}
\label{SM}
{D}(N_s,N_t)=2N_t\log(N_t)-2(N_t+1)\log(N_t+1)-(N_s+N_t+2N_sN_t)\log(N_s+N_t+2N_sN_t)+\\
-\frac{N_t(N_t+1)}{1+N_s+N_t+2N_sN_t}
\log\left(\frac{N_t(N_t+1)}{1+N_s+N_t+2N_sN_t}\right)
+(1+N_s+N_t+2N_sN_t)\log(1+N_s+N_t+2N_sN_t)\\
+\frac{N_s+2N_sN_t+(1+N_t)^2}{1+N_s+N_t+2N_sN_t}
\log\left(\frac{N_s+2N_sN_t+(1+N_t)^2}{1+N_s+N_t+2N_sN_t}\right)
\end{aligned}
\end{equation}
expressed as a function of the effective parameters $N_s$ and $N_t$. By 
using  Eqs. (\ref{eq:NsNt}), one can easily obtain the discord 
as a function of the physical parameters $r$, $\gamma$ and
$\eta$.\\
\subsection{Monte Carlo evaluation of uncertainties.}
In our experiment, $M_q$ measurement outcomes are recorded for each one
of the four quadratures. The quadratures $Q^{(1)}$ and $Q^{(4)}$ form a
set of $2M_q$ squeezed quadratures measurements, as well as $Q^{(1)}$
and $Q^{(4)}$ form a set of $2M_q$ anti-squeezed quadratures
measurements. From those two sets of experimental data, we compute the
variances $\sigma^2(Q_{\rm sq})$ and $\sigma ^2(Q_{\rm asq})$. Assuming
that those estimated variances follow a Gaussian distribution, the
variance of their estimation is given by
$\textrm{Var}\left(\sigma^2(Q_\mathrm{sq/asq})\right){=}2\sigma^4(Q_\mathrm{sq/asq})/(2
M_q)$.

The expressions for $\sigma^2(Q_{\rm sq})$ and $\sigma ^2(Q_{\rm asq})$
are then inverted  to obtain $N_s$ and $N_t$ as function of $\sigma
^2(Q_{\rm sq})$ and $\sigma ^2(Q_{\rm asq})$. The mean value and the
associated uncertainties are determined by a Monte Carlo simulation of
$10^6$ experiments. For each experiment, the values $\tilde{\sigma}
^2(Q_{\rm sq})$ and $\tilde{\sigma} ^2(Q_{\rm asq})$ of the squeezed and
anti-squeezed variances are randomly chosen from two Gaussian
distributions respectively of mean values $\sigma ^2(Q_{\rm sq})$ and
$\sigma ^2(Q_{\rm asq})$, and variances
$\textrm{Var}\left(\sigma^2(Q_\mathrm{sq})\right)$ and
$\textrm{Var}\left(\sigma^2(Q_\mathrm{asq})\right)$. The values of
$\tilde{N}_s$ and $\tilde{N}_t$ are then computed using those random
values.  The experimental estimate $N_s^{\rm inv} $ and $N_t^{\rm inv}$
are finally obtained by taking the mean of the $10^6$ values of
$\tilde{N}_s$ and $\tilde{N}_t$, whereas their uncertainties
$\sigma^2(N_s^{\rm inv} )$ and $\sigma^2(N_t^{\rm inv} )$ are obtained
by computing the variance of the $10^6$ values of $\tilde{N}_s$ and
$\tilde{N}_t$.

These values can be used in the expression for discord \eqref{SM} to
calculate its value $D^{\rm inv}$, and the uncertainty $\sigma^2(D^{\rm
inv})$ by using a similar MonteCarlo method.
\end{document}